\documentclass[%
 aps,prx,
 amsmath,amssymb,
reprint,
twocolumn
]{revtex4-1}

\usepackage{float}
\usepackage[hypertexnames=false]{hyperref}
\hypersetup{
    colorlinks = true,
    allcolors = {blue}
}
\usepackage[all]{hypcap}   
\usepackage{enumitem}
\usepackage{graphicx}
\begin{document}

\title{Ice is Born in Low-Mobility Regions of Supercooled Liquid Water}

\author{Martin Fitzner}
\affiliation{Thomas Young Centre, London Centre for Nanotechnology and Department of Physics and Astronomy, University College London, Gower Street London WC1E 6BT, United Kingdom}

\author{Gabriele C. Sosso}
\affiliation{Department of Chemistry and Centre for Scientific Computing, University of Warwick, Gibbet Hill Road, Coventry CV4 7AL, United Kingdom}

\author{Stephen J. Cox}
\altaffiliation{Present address: Department of Chemistry, University of Cambridge, Lensfield Rd, Cambridge CB2 1EW, United Kingdom}
\affiliation{Thomas Young Centre, London Centre for Nanotechnology and Department of Physics and Astronomy, University College London, Gower Street London WC1E 6BT, United Kingdom}

\author{Angelos Michaelides}
\altaffiliation{To whom correspondence should be addressed. E-mail: angelos.michaelides@ucl.ac.uk}
\affiliation{Thomas Young Centre, London Centre for Nanotechnology and Department of Physics and Astronomy, University College London, Gower Street London WC1E 6BT, United Kingdom}
\date{\today}

\begin{abstract}
When an ice crystal is born from liquid water two key changes occur: (i) the molecules order; and (ii) the mobility of the molecules drops as they adopt their lattice positions. 
Most research on ice nucleation (and crystallization in general) has focused on understanding the former with less attention paid to the latter. 
However, supercooled water exhibits fascinating and complex dynamical behavior, most notably dynamical heterogeneity (DH), a phenomenon where spatially separated domains of relatively mobile and immobile particles coexist. 
Strikingly, the microscopic connection between the DH of water and the nucleation of ice has yet to be unraveled directly at the molecular level.
Here we tackle this issue via computer simulations which reveal that: (i) ice nucleation occurs in low-mobility regions of the liquid; (ii) there is a dynamical incubation period in which the mobility of the molecules drops prior to any ice-like ordering; and (iii) ice-like clusters cause arrested dynamics in surrounding water molecules.
With this we establish a clear connection between dynamics and nucleation.
We anticipate that our findings will pave the way for the examination of the role of dynamical heterogeneities in heterogeneous and solution-based nucleation.
\end{abstract}

\maketitle

The freezing of water is one of the most ubiquitous phase transitions on earth, shaping many processes from intracellular freezing~\cite{mazur1970cryobiology} to cloud formation~\cite{vergara2018strong}, yet there are major gaps in our molecular-level understanding of ice nucleation~\cite{bartels-rausch_chemistry:_2013}. Considerable effort has gone into understanding supercooled water and the microscopic details of ice nucleation~\cite{sellberg_ultrafast_2014,malkin2015stacking,kim2017maxima,palmer2018advances}. In particular, computer simulations have proved to be of great importance in understanding ice nucleation recently. They have for instance shown that the nucleation of hexagonal ice I$_\mathrm{h}$ proceeds through stacking-disordered ice I$_\mathrm{sd}$~\cite{moore_structural_2011,li_homogeneous_2011,li2013ice,haji-akbari_direct_2015} and that this polytype is entropically stabilized at cluster sizes relevant to nucleation~\cite{lupi_Nature2017}. While this has greatly advanced our understanding of the structural transformation during the nucleation process, nucleation is likely also influenced by the particle mobility, i.e. the dynamics of the liquid. The liquid dynamics during or shortly before ice nucleation has however not been directly studied or assessed in experiment. This is particularly striking since water exhibits dynamical heterogeneity (DH)~\cite{ediger2000spatially}, the phenomenon of coexisting spatially extended domains of very different mobility. 

\begin{figure}[t!]
\begin{centering}
\centerline{\includegraphics[width=9.2cm]{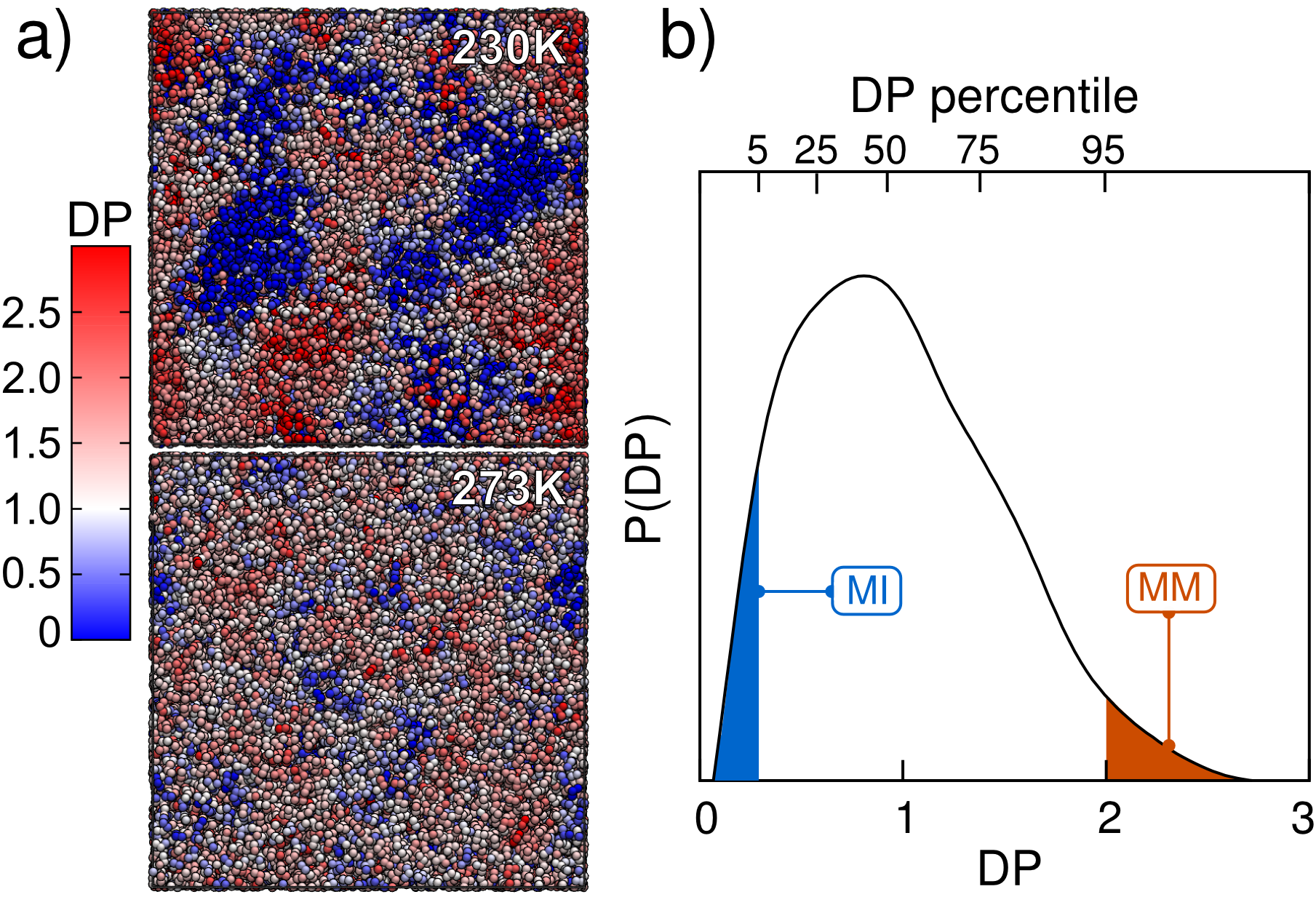}}
\par\end{centering}
\protect
\caption{Dynamical heterogeneity in supercooled liquid water with the TIP4P/Ice model.
a) Spatial distribution of the dynamical propensity (DP) at 230 and 273~K. Molecules (only oxygens shown) are colored according to the scale on the left. b) Probability density distribution of the DP at 230~K. Blue and red shaded regions highlight the 5\% of water molecules labeled as most immobile (MI) and most mobile (MM).}
\label{FIG_1}
\end{figure}
\begin{figure*}[t!]
\begin{centering}
\centerline{\includegraphics[width=18cm]{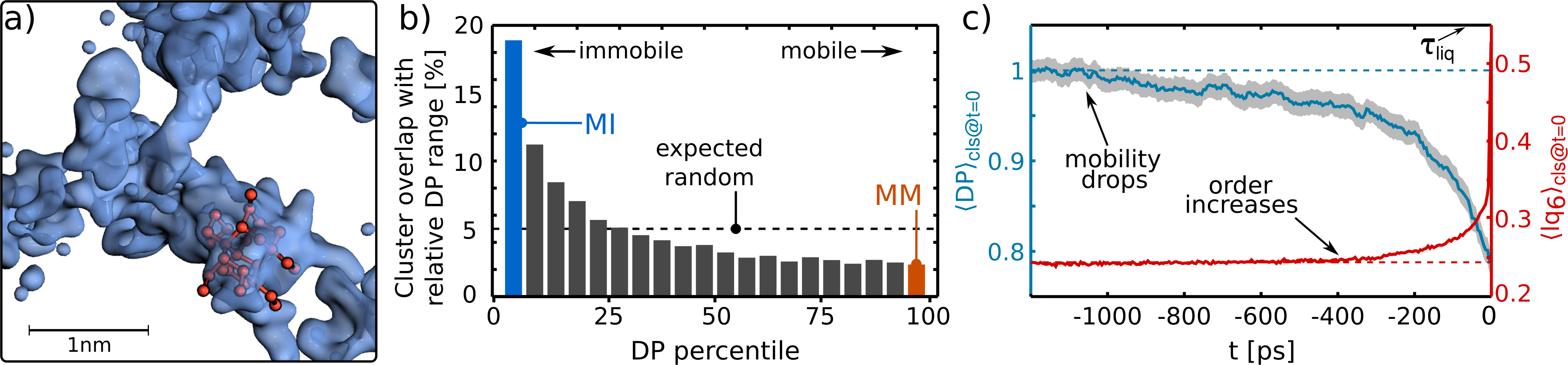}}
\par\end{centering}
\protect
\caption{Connection between dynamical heterogeneity (DH) and pre-critical cluster formation in the TIP4P/Ice model.
a) Representative snapshot of a spontaneously formed cluster (red bonds and spheres) immersed in the MI region (transparent blue surface representation).
b) Average overlap between the molecules in the largest ice-like cluster and molecules in the relative DP range. Each bar corresponds to a 5\% fraction of (sorted) DP values, i.e. the first / last bar corresponds to the MI / MM regions. The expected overlap if clusters were uncorrelated with the DP would be 5\% (indicated by the dashed line).
c) Average evolution of the mobility (DP) and crystallinity (lq$_6$) for molecules in a cluster before its first time of assembly (taken to be $t=0$). Dashed lines indicate the mean values of DP and lq$_6$ of the liquid. $\tau_\mathrm{liq}$ is the structural relaxation time. Shaded regions indicate 95\% confidence intervals. All data was obtained with the TIP4P/Ice water model at 240~K.}
\label{FIG_2}
\end{figure*}
At the molecular scale, there are several studies that found structural differences between very mobile and very immobile liquid regions. Sciortino \textit{et al.} have shown that increasing local coordination leads to higher diffusion of molecules~\cite{sciortino_effect_1991}. Molinero and coworkers have seen that ice originates from four-coordinated regions both in the pure liquid~\cite{moore_structural_2011} and in salt-water solutions~\cite{bullock2013low}, as well as a growing correlation length of four coordinated water patches~\cite{moore2009growing}. Regarding bond-orientational order, Tanaka and coworkers have shown that less mobile regions have a higher degree of tetrahedrality~\cite{russo2018water}, consistent with other studies~\cite{errington_cooperative_2002,matsumoto_molecular_2002}, and they also identified that 5-membered rings could act as locally favored structures~\cite{russo_understanding_2014}. Recently they have also shown that the emergence of DH can be described by a two-state picture of ordered and disordered regions~\cite{shi2018origin}.

The pre-ordering in less mobile domains could be seen as an indication that nucleation will be preferred in these regions~\cite{golde2016correlation}. On the other hand Mazza \textit{et al.}~\cite{mazza2006relation} found that rotational and translational heterogeneities correlate in water, and thus, the transformative motions of the rearrangements necessary for crystallization could be hindered in strongly immobile regions. Indeed, it has been argued for metallic liquids~\cite{zhang2018spatially} that the enhanced mobility near the surface of nanowires can explain surface-induced crystallization rather than the traditional view of a heterogeneous reduction of the nucleation barrier. Mobile molecules would have the potential for a more effective exploration of phase-space and therefore the ability to more readily perform collective self-assembly. Overall, one can find arguments for both immobile and mobile regions being preferential domains for nucleation. But whether there is any preference at all is not established to date for water, or indeed any other supercooled liquid.
\begin{figure*}[t!]
\begin{centering}
\centerline{\includegraphics[width=17cm]{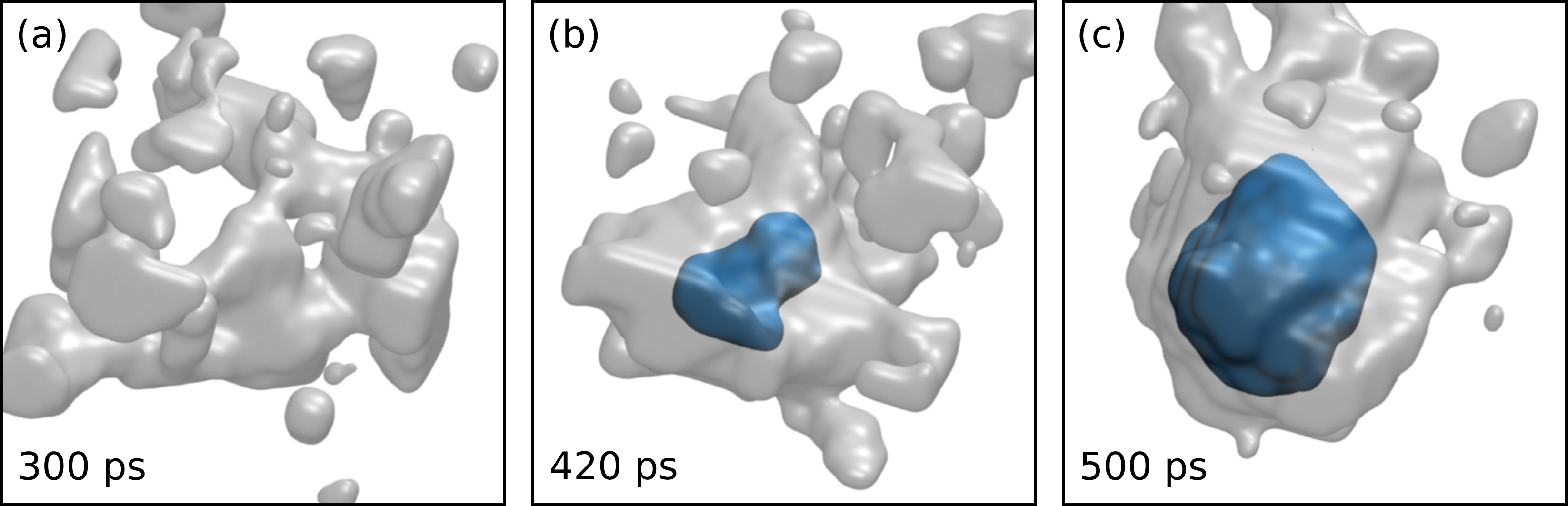}}
\par\end{centering}
\protect
  \caption{Ice nucleation occurs in relatively immobile domains of supercooled water. Time evolution of the coarse-grained immobility $\mathcal{I}(\mathbf{r})$ (translucent silver) and crystallinity $\mathcal{Q}(\mathbf{r})$ (opaque blue) fields, from a trajectory harvested by TPS with the mW model. (a) Prior to nucleation we see large immobile domains and an absence of crystalline order. During nucleation, the ice nucleus forms (b) and grows (c) within the immobile domain. The ice cluster in snapshots b) and c) comprised of 83  and 296 molecules respectively. The diameter of the ice-like region in c) is approximately 3.4~nm.
}
\label{FIG_3}
\end{figure*}

Here, we fill this gap by performing computer simulations of supercooled liquid water and ice nucleation which show that: (i) ice nucleation occurs in low-mobility regions of the liquid; (ii) there is a dynamical incubation period in which the mobility of the molecules drops prior to ice-like ordering; and (iii) ice-like clusters cause arrested dynamics in surrounding water molecules. 
The remainder of this manuscript is structured as follows: First, we uncover and characterize DH in the liquid. We then study the connection of DH with pre-critical ice clusters in an atomistic water model. This is followed by an analysis of ice nucleation trajectories obtained from enhanced sampling with a coarse-grained water model. The results are then corroborated by studying the structural differences between the most mobile and most immobile liquid regions. Lastly, we conclude and discuss some of the implications of our findings.
\section*{Uncovering Dynamical Heterogeneity}
We begin by characterizing the extent of DH in supercooled liquid water represented by the atomistic TIP4P/Ice~\cite{abascal2005potential} model. To this end we perform molecular dynamics (MD) simulations in a homogeneous water system in the temperature range 230~K to 273~K, utilizing iso-configurational analysis (ISOCA)~\cite{widmer-cooper_study_2007,sosso_dynamical_2014}. This technique allows us to obtain spatially resolved maps of DH. We quantify the tendency of each molecule to move with a \textit{dynamical propensity} (DP):
\begin{equation}
\mathrm{DP}_i = \left<\frac{\lVert\mathbf{r}_i(t_0)-\mathbf{r}_i(0)\rVert^2}{\mathrm{MSD}}\right>_\mathrm{ISO}
\label{dpeq}
\end{equation}
where $\mathbf{r}_i(t)$ is the position vector of molecule $i$ at time $t$, $t_0$ is the time of maximum heterogeneity (see definition in Methods) and $\mathrm{MSD}$ is the mean square displacement of all oxygen atoms. In this approach we average the outcome over many trajectories that start from the same initial configuration but with a different set of random velocities, indicated by the notation $\langle \dots \rangle_{\mathrm{ISO}}$. By doing this we are able to evaluate the effect of structure alone on DH. In Fig.~\ref{FIG_1}a we show two snapshots of initial configurations used for the ISOCA at 230 and 273~K, where each oxygen atom is colored according to its $\mathrm{DP}$. This choice of temperatures is to illustrate the maximum difference in the extent of DH as well as the relevance of strong supercoolings like 230~K in homogeneous nucleation of ice~\cite{nilsson2015structural,lupi_Nature2017,haji-akbari_direct_2015}. It can be seen that spatially localized domains of relatively immobile (blue) and mobile (red) particles emerge and that the spatial aggregation of the domains differs drastically. As reported in Fig.~\ref{FIG_1}b, the probability density distribution of the $\mathrm{DP}$ at 230~K is rather broad with a factor of 30 between the mobility of particles at opposite tails. From the distribution of the DP we select the top and bottom 5\% and label the corresponding molecules as most mobile (MM) and most immobile (MI) regions for further analysis. We show in the supporting information (SI) that the choice of this threshold has no major influence on our results.

\section*{Dynamics of Pre-Critical Fluctuations}
The simulation of nucleation with atomistic water models currently remains a challenge, coming at enormous computational cost~\cite{haji-akbari_direct_2015}. Hence,
as a first step to understand the connection between DH and ice nucleation, we focus on pre-critical clusters, i.e. the ice-like clusters that form via frequent thermal fluctuations and thus are readily probed by unbiased MD~\cite{fitzner2017pre}. The key results of our simulations with the TIP4P/Ice model are reported in Fig.~\ref{FIG_2}, where we find a strong tendency for the pre-critical ice nuclei to form within MI domains, rather than the MM domains. To quantify their preference to form in the immobile regions we have split the whole range of (sorted) DP values into 20 equal sections (i.e. each corresponding to 5\% of the whole range). This means that the molecules in the first/last DP section are the same molecules as the ones in the MI/MM regions. In Fig.~\ref{FIG_2}b we plot now the average overlap of molecules in the biggest ice-like cluster with these DP sections. The expected overlap in the absence of any correlation with the DP would be 5\%. We clearly see that there is a strong preference for pre-critical ice clusters to belong to the lower DP sections (i.e. more immobile molecules). In addition, we find that the formation of pre-critical clusters is suppressed in the MM domains as the overlap values there are below the base-line of 5\%. Considering the fact that nucleation is stochastic in nature, this is strong evidence for the connection between immobile and ice-forming regions.

Having shown that pre-critical ice clusters strongly overlap with the immobile regions from their early stages we now study the temporal connection between immobility and clusters \textit{before} their first time of assembly. In Fig.~\ref{FIG_2}c we show the average value of the DP and the crystallinity parameter lq$_6$~\cite{li_homogeneous_2011} of molecules that belong to a cluster at its first time of assembly (taken to be at $t = 0$). It can be seen that the mobility drops at $\sim -1000$~ps, which is significant compared to the structural relaxation time of the liquid: $\tau_\mathrm{liq} \approx 68$~ps at 240~K (see Methods). In addition this drop occurs earlier and is much less abrupt than the change in structure, which can be associated with the rise of the lq$_6$ order parameter at about 400~ps before the assembly. This finding is crucial and it confirms that immobility on average precedes ice-cluster formation by a significant timespan. This can be thought of as a \textit{dynamical incubation} period in which the dynamics of the molecules changes prior to the structural change towards ice. While this mechanism is not necessarily orthogonal to the commonly applied reasoning of purely structural ordering, our results suggest that arguing in terms of a process that involves distinct dynamical and subsequent structural steps is a viable route for a better description of nucleation.

\section*{Connection Between Nucleation and Dynamics}
The results reported above point strongly towards an interplay between structural motifs pertinent to nucleation and the dynamics of the system. Unbiased MD simulations however, cannot directly sample nucleation events, except under extreme conditions close to the homogeneous nucleation temperature \cite{moore_structural_2011}. In this section, we report results from transition path sampling (TPS)~\cite{bolhuis_transition_2002} simulations that allow us to sample many nucleation events at reasonably high temperature. Specifically, we study a system of water molecules represented by the coarse-grained mW model~\cite{molinero_water_2009} at 235~K. In the case of the brute force simulations, we employed an ISOCA to quantify mobility. However, in the case of TPS we harvest many more (7500) reactive trajectories, making ISOCA for each frame of each trajectory impractical. To fully exploit the statistical sampling provided by TPS, we therefore employ the \emph{enduring displacement}~\cite{speck_constrained_2012,limmer_theory_2014} formalism, which permits on-the-fly calculation of each particle's mobility. Using this approach has the added benefit of allowing us to validate the robustness of our previous results we show that there exists a correspondence between quantifying the mobility with enduring displacements and the DP method in the SI).

In order to identify regions of space as either ice-like/liquid-like and immobile/mobile (and the boundaries separating them), we introduce the coarse-grained crystallinity field $\mathcal{Q}(\mathbf{r})$, and immobility field $\mathcal{I}(\mathbf{r})$. In brief, $\mathcal{Q}(\mathbf{r})$ is obtained by smearing each ice-like molecule's position with a positive, normalized Gaussian, and each liquid-like molecule with a negative Gaussian. In regions that are predominantly ice-like, $\mathcal{Q}(\mathbf{r})$ will take values close to the density of ice, whereas in regions that are predominantly liquid-like, it will take values close to the negative density of liquid water. Boundaries separating ice-like and liquid-like regions are defined by surfaces with $\mathcal{Q}(\mathbf{r})=0$. In the SI, we show that $\mathcal{Q}(\mathbf{r})$ performs well at identifying regions as ice-like that are consistent with our intuitive understanding, while simultaneously neglecting small fluctuations. In a similar fashion, $\mathcal{I}(\mathbf{r})$ is obtained by smearing immobile particles with a positive Gaussian, and mobile particles with a negative Gaussian.
\begin{figure}[t!]
\begin{centering}
\centerline{\includegraphics[width=8.4cm]{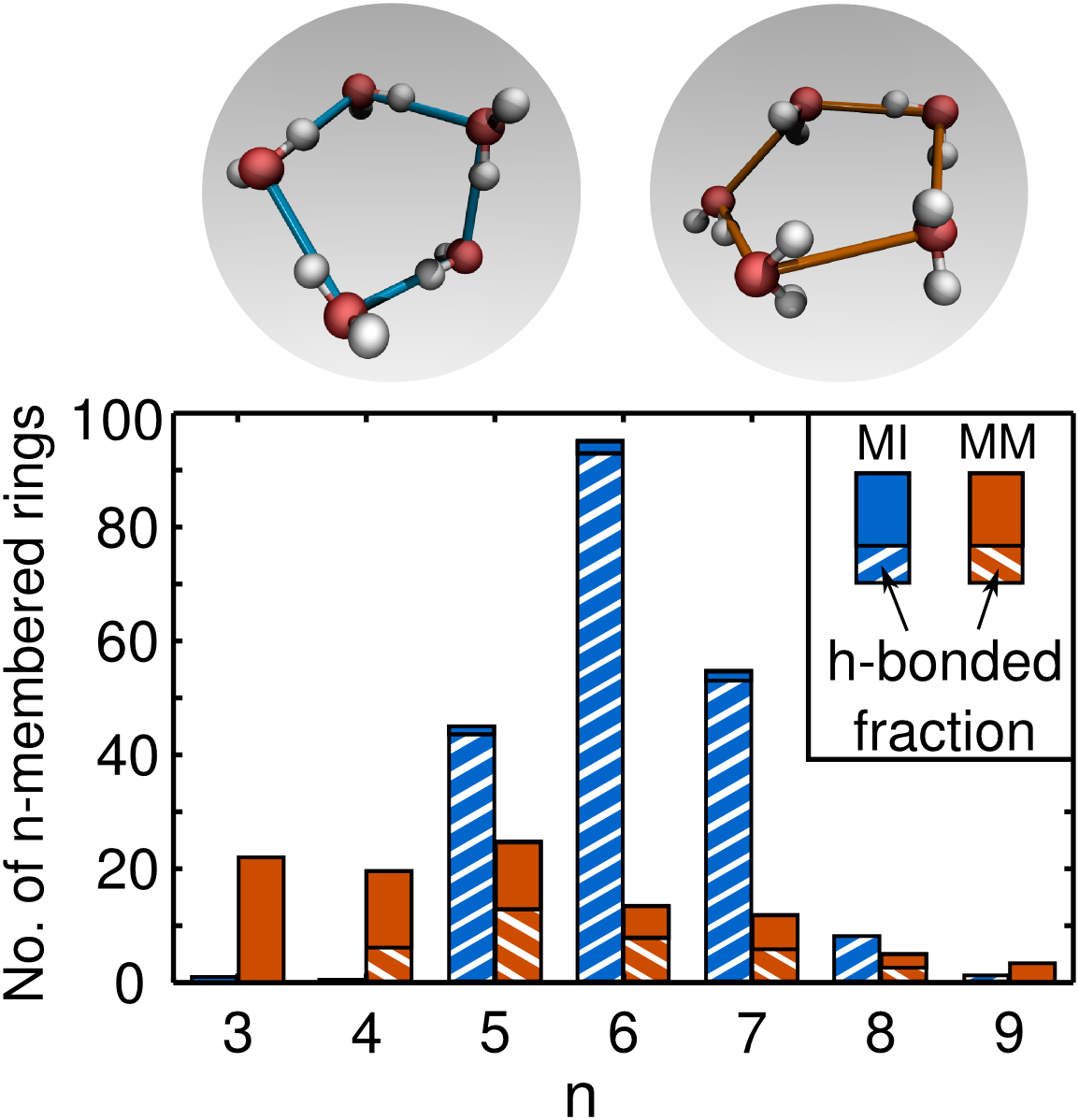}}
\par\end{centering}
\protect
\caption{Structural differences in regions of adverse mobility in TIP4/Ice water.
Number of $n$-membered primitive rings within the respective domain at 230~K. The dashed portions of the bars represent the fraction of those rings fully connected by H-bonds. The top insets show an example for a fully and non-fully H-bonded 5-membered ring, where solid lines between oxygens are a guide to the eye and do not imply H-bonds.}
\label{FIG_4}
\end{figure}

In Fig.~\ref{FIG_3}a we show a snapshot of $\mathcal{Q}(\mathbf{r})$ and $\mathcal{I}(\mathbf{r})$ prior to nucleation (only regions $\mathcal{Q}(\mathbf{r}) > 0$ and $\mathcal{I}(\mathbf{r}) > 0$ are shown) from a typical trajectory harvested from TPS. While there are no regions identified as ice-like, we do see large immobile domains in the supercooled liquid. Figs.~\ref{FIG_3}b,c show similar snapshots after the onset of nucleation. Crucially, and consistent with our findings for the pre-critical nuclei, it is clear that the ice nucleus forms within an immobile domain. In addition to this trajectory, we calculated statistical quantities characterizing the whole ensemble of TPS trajectories. These are discussed in the SI and show that the behavior identified in Fig.~\ref{FIG_3} is indeed typical for all nucleation trajectories.

Overall, the results obtained from TPS support and extend our observations from the unbiased simulations and strengthen our conclusion since we now also have insight: (i) for larger clusters; (ii) for a large ensemble of nucleation trajectories; and (iii) with a different method of classifying immobility. In Fig.~\ref{FIG_3}b,c it can also be seen that the surrounding immobile region is larger than the ice-like region, which suggests that ice-clusters further slow down their surroundings. Indeed, we also find for the atomistic model (see SI) that water molecules within approximately 2.5 hydration shells have arrested dynamics as their DP values are substantially lower than in the bulk. This is relevant to crystal growth and theoretical modeling as the liquid molecules in direct vicinity are significantly less mobile than those in the bulk.

\section*{Structural Hallmarks of Nucleating Regions}
Given the presented evidence that shows that ice nucleation occurs in relatively immobile regions rather than mobile ones, it is interesting to investigate the structural differences between these two domains. As mentioned in the introduction, there have been a number of works already highlighting different kinds of pre-ordering, i.e. regarding tetrahedrality~\cite{sciortino_effect_1991,matsumoto_molecular_2002} or coordination number~\cite{sciortino_effect_1991,moore_structural_2011}. We will add to this by analyzing the distribution of primitive rings (i.e. not divisible into smaller ones) in the regions of extreme mobilities (MM and MI as defined in figure~\ref{FIG_1}b) for the atomistic water model. As can be seen from Fig.~\ref{FIG_4} the MI regions have a rings distribution strongly peaked around $6\pm1$ members while the distribution for the MM domains is very broad. Moreover, the amount of entirely H-bonded rings is substantially higher in the MI region. Indeed, if we were to regard hydrogen bonds between members as a necessary criterion for being a ring (as is done in other literature~\cite{haji-akbari_direct_2015,kusalik_rings2}), the MM region would be almost free of rings. In particular, an abundance of 6$\pm1$-membered hydrogen bonded rings can be regarded as the key structural characteristic of the MI domains in the liquid. 

These results add to the findings of Haji-Akbari and Debenedetti~\cite{haji-akbari_direct_2015}, who showed that the nucleating ice nucleus exhibits a similar rings distribution, and Pirzadeh \textit{et al.} who highlight the presence of $6\pm1$-membered rings near growing ice surfaces~\cite{kusalik_rings2}. We stress here, however, that in the liquid snapshots we investigated for the rings analysis, we find only negligible amounts of actual ice (according to different criteria, for details see the SI). Since the majority of 6-membered rings in the MI domains can be seen as ice-like if regarded in isolation, this means that it is the relative orientation between rings that is different from the crystal, and thus the missing ingredient in forming ice. Because this happens in the MI region, which has a reduced diffusivity compared to other regions, we can speculate that the mechanism giving rise to the initial formation of ice-like clusters in the liquid is collective in nature (a similar argument based on density changes was made by Errington \textit{et al.}~\cite{errington_cooperative_2002}). This would be consistent with a picture of reorienting rings rather than a picture of single-particle attachments via diffusive motion.

\section*{Discussion \& Conclusions}
Our results have established a clear link between dynamical heterogeneity and ice nucleation, suggesting that the complex nature of dynamics should not be overlooked in theoretical descriptions of nucleation. We have shown that liquid molecules in the vicinity of a nucleus are slowed down significantly, which implies that their diffusivity (connected to the attachment rate) is reduced compared to bulk molecules, an aspect neglected by classical nucleation theory. 

We verified that our results on the rings distribution also hold for the coarse-grained mW model~\cite{molinero_water_2009} even though for mW the extent of DH is much smaller (see SI). This means that the characteristic features identified in our study are not sensitive to the specific hydrogen bond parametrization, but rather caused by the tetrahedral order inherent in the modeled material. Thus, our findings may be of relevance not just to water but to a much broader range of materials, evidently ones with tetrahedral order (such as group IV elements and silica). More broadly, it remains to be seen if the immobility of pre-crystalline structures is connected to nucleation in non-tetrahedral liquids in the same manner since the population of e.g. rings will be material-specific. However, based on our findings we can suggest that it is the correspondence between immobile and crystalline topological features (such as rings) that connects immobile regions with nucleation.

The connection of DH and nucleation in water could be probed experimentally by investigating heavier water molecules as their mobility might be different. 
It is for instance established that liquid D$_2$O has a higher nucleation rate than H$_2$O~\cite{stockel2005rates}. However this cannot be taken as direct evidence in support of our observation as the change in the hydrogen bonding induced by nuclear quantum effects~\cite{ceriotti2016nuclear} potentially influences the nucleation rate too. 
A more rigorous experimental validation of our findings would be the nucleation rate comparison for H$_2 {}^{18}$O and ordinary water, which to the best of our knowledge has not been achieved. If one of the two liquids is more/less mobile (diffusive) our results suggest a decreased/enhanced nucleation rate. 

Our findings are likely of relevance to heterogeneous nucleation and nucleation from solution. Generally, previous work has focused on investigating the structural or templating role of nucleating agents~\cite{fitzner_many_2015,lupi_heterogeneous_2014,sosso_crystal_2016,qiu2017ice}. However, an impurity or substrate is bound to impact the dynamics of the supercooled liquid in its vicinity, possibly leading to a novel mechanism of heterogeneous nucleation. For the example of water freezing, ice nucleation on hydrophobic surfaces (basically incapable of structuring the water network to a major extent) has been reported~\cite{lupi_heterogeneous_2014,fitzner_many_2015} as well as intriguing alternating hydrophilic-hydrophobic patterns in the prominent ice nucleating bacteria \textit{pseudomonas syringae}~\cite{pandey_ice-nucleating_2016} and a nucleation enhancement by soluble molecules~\cite{mochizuki2017promotion} that could be connected to the liquid dynamics. Moreover, for nucleation from solution it is well known that different solutes change the nucleation mechanisms, i.e. in the case of urea~\cite{salvalaglio2015urea}. Understanding how solutes change the dynamics and impact the formation of amorphous precursors could shed light on this issue. As such, we hope that this work will push the community to take into account the role of dynamics and particularly of DH in connection with crystal nucleation and growth.

In conclusion, we have shown that ice nuclei originate within immobile regions of the supercooled liquid and that there is a dynamical incubation period in which the mobility of particles drops prior to any structural change. Additionally, the presence of an ice crystallite causes arrested dynamics in water molecules that surround it and the distribution of rings can be seen as the structural hallmarks of DH in water. This connection between dynamics and structure provides a novel perspective on the physics of nucleation.

\section*{Methods}
\subsection*{Molecular Dynamics Simulations}
We mainly study the DH of a system containing 10,000 water molecules, represented by the TIP4P/Ice model~\cite{abascal2005potential}. All our MD simulations are performed with the LAMMPS code~\cite{plimpton1995fast}, integrating the equations of motion with a 2~fs time-step and using a ten-fold Nos\'{e}-Hoover chain~\cite{martyna1992nose} with a relaxation time of 200~fs to control temperature. We employ a cubic simulation box with 3-dimensional periodic boundary conditions and approximate volume of 68$\times$68$\times$68~\r{A}$^3$. Static bonds and angles have been constrained with the SHAKE algorithm~\cite{ryckaert1977numerical}. To avoid quenching effects upon generating starting configurations at different temperatures we performed (after 10~ns equilibration at melting temperature) a 0.5~K/ns cooling ramp in the NPT ensemble (ten-fold Nos\'{e}-Hoover chain barostat with relaxation time of 2~ps). At 273, 260, 250, 240, 230, 220 and 210~K we save configurations. Those are propagated for 10~ns in the NVT ensemble at equilibrium volume to calculate dynamical properties as well as drawing 5 snapshots for each temperature that are apart at least 1~ns to use for the ISOCA. 

\subsection*{Work Flow to Characterize the Liquid Dynamics}
In order to characterize the liquid dynamics appropriate length and time scales have to be chosen, which is achieved by the following procedure:
\begin{enumerate}
	\item From the NVT simulation of the system with $N = $ 10,000 molecules at the target temperature we obtain the oxygen-oxygen radial distribution function $g_\mathrm{OO}(r) = \left\langle \frac{1}{2\pi r^2 N \rho}\sum_{i=1}^{N} \sum_{j > i}^{N} \delta\left(r - \Vert \mathbf{r}_i-\mathbf{r}_j \Vert\right) \right\rangle$ where the sums consider $N$ oxygen atoms and their positions $\mathbf{r}_{i/j}$, $\rho$ is the liquid density and the average is over all trajectory frames.
    \item We calculate the isotropic structure factor $S(q) = 1 + \frac{4\pi\rho}{q} \int_0^\infty \mathrm{d}r\,r\,\sin(rq)\,\left[g_\mathrm{OO}(r)-1\right]$ and define $q_0$ as the value where $S(q)$ has its first peak, with $q$ being a reciprocal length.
    \item For this $q_0$ we calculate the quantity $\Phi(\mathbf{q},t) = \frac{1}{N}\sum_{j=1}^N\exp(i\mathbf{q}\cdot[\mathbf{r}_j(t)-\mathbf{r}_j(0)])$
    \item Via $\Phi(\mathbf{q},t)$ we obtain the self-intermediate scattering function $F(\mathbf{q},t) = \left<\Phi(\mathbf{q},t)\right>$ and the dynamical susceptibility $\chi_4(\mathbf{q},t) = N\left[\left<\left|\Phi(\mathbf{q},t)\right|^2\right>-\left<\Phi(\mathbf{q},t)\right>^2\right]$. Isotropic averages taken over 200 independent directions according to $F(q_0,t) = \left<F(\mathbf{q},t)\right>_{\lVert\mathbf{q}\rVert=q_0}$ are evaluated.
    \item The \textit{time of maximum heterogeneity} $t_0$ is taken as the time where $\chi_4(q_0,t)$ has its maximum, i.e. where the movements at the nearest-neighbor range are most heterogeneous.
\end{enumerate}
The resulting values for $q_0$ and $t_0$ for all temperatures can be found in Table~\ref{TAB_SI_1} together with the structural relaxation time $\tau_\mathrm{liq}$ that was obtained as $\alpha$-relaxation value from $F(q,t)$.
\begin{table}[ht]
\begin{tabular}{l|ccccccc}
\hline
                         & 273~K & 260~K & 250~K & 240~K & 230~K  & 220~K & 210~K \\ \hline
$q_0$ [\r{A}$^{-1}$]     & 2.01  & 1.96  & 1.91  & 1.84  & 1.80   & 1.77  & 1.76  \\ \hline
$t_0$ [ps]               & 5     & 11    & 27    & 115   & 620    & -     & -     \\ \hline
$\tau_\mathrm{liq}$ [ps] & 2     & 6     & 14    & 68    & 356    & -     & -     \\ \hline
\end{tabular}
\caption{Overview of length ($q_0$) and time ($t_0$) scales used to characterize dynamical heterogeneity and structural relaxation time $\tau_\mathrm{liq}$ at different temperatures with the TIP4P/Ice model. Because of the computational cost we did not consider 220~K and 210~K for the rest of the study.}
\label{TAB_SI_1}
\end{table}

\subsection*{Analyzing the Connection Between Water Structure and Dynamics}
Ice-like molecules were detected using an order parameter (lq$_6$) according to Li \textit{et al.}~\cite{li_homogeneous_2011} as implemented in PLUMED2~\cite{tribello2014plumed,tribello_analyzing_2017}. First we compute for each molecule $i$ the quantity:
\begin{equation}
	q_{lm}(i) = \frac{1}{N_b(i)}\sum_{k=1}^{N_b(i)}Y_{lm}(\theta_{ik},\phi_{ik})
\end{equation}
where the sum goes over the $N_b(i)$ neighbors of molecule $i$, $Y_{lm}$ are spherical harmonics and $\theta_{ik}$ and $\phi_{ik}$ are the relative orientational angles between the molecule $i$ and $k$. We then compute this quantity for all possible values of $m$ and store them in a vector $\vec{q}_l(i)$ with $2l + 1$ components. Finally we calculate values $\mathrm{lq}_l$ according to:
\begin{equation}
\textrm{lq}_l(i) = \frac{1}{N_b(i)}\sum_{k=1}^{N_b(i)}\frac{\vec{q}_l(i)\cdot\vec{q}_l(k)}{\left|\vec{q}_l(i)\right|\cdot\left|\vec{q}_l(k)\right|}
\end{equation}
For the particular choice of $l = 6$ we classify a molecule with a value of lq$_6 > 0.5$ as ice-like, otherwise as liquid. Ice-like molecules are then grouped together if they are within 3.4\r{A} of each other, and we call the resulting entities ice-like clusters.

For all snapshots (5 for each of the 5 different temperatures) we calculate the dynamical propensity (DP, see Eq.~\ref{dpeq}) by performing 200 independent MD runs in the NVT ensemble, starting with randomized velocities. As length of these production runs we choose the time $t_0$. For our purpose it is sufficient to consider oxygens only when calculating the DP. The latter is then used to label the most mobile (MM) and most immobile (MI) molecules as the top/bottom 5\% of DP values in the respective snapshot. The structural properties of these regions are then established by calculating the number of primitive $n$-membered rings within them (utilizing the R.I.N.G.S. code~\cite{le2010ring}), as well as computing distributions of common order parameters such as $q_\mathrm{tetra}$ or topological patterns such as cages. All results reported in the text are averages over the results for the 5 snapshots per temperature.

The results reported in Fig.~\ref{FIG_2}b,c were calculated from a TIP4P/Ice trajectory at 240~K. We define as the time of first assembly of a cluster the frame that for the first time has the biggest ice-like cluster comprised of molecules that have not been part of the biggest ice-like cluster in the 75~ps before that frame. This is slightly larger than the structural relaxation time of 68~ps at that temperature and larger or smaller choices did not qualitatively alter the findings. 
The 95\% confidence intervals for the DP evolution result from point-wise estimates from 10$^6$-fold bootstrap resampling of all relevant clusters, the corresponding intervals for the lq$_6$ curve are smaller than the line width plotted.

\subsection*{Transition Path Sampling Simulations}
Transition path sampling~\cite{bolhuis_transition_2002} (TPS) was performed using an in-house code interfacing with LAMMPS~\cite{plimpton1995fast}. The system comprised of 4000 mW molecules and a pressure of 1 atm was maintained using a barostat with a damping constant of 5~ps and a timestep of 10~fs. Langevin dynamics was used to maintain a temperature of 235~K with a damping constant of 10~ps.

To define whether or not a trajectory was reactive or not, we used the size of the largest ice-like cluster $N_\mathrm{cls}$ as defined by Li \emph{et al}.~\cite{li_homogeneous_2011}. This means, we additionally include each ice-like molecule's nearest neighbor into the cluster to effectively include a surface contribution. The system was considered to be liquid if $N_\mathrm{cls} < $ 50 and ice if $N_\mathrm{cls} > $ 800. An initial reactive trajectory was generated by unbiased simulation at 205~K and the velocities were rescaled by a factor $\sqrt{\frac{235}{205}}$. An equilibration of 1000 TPS moves was then performed with a 2:1 ratio of {\it shooting} to {\it shifting} moves. We made use of the {\it one-way shooting} algorithm~\cite{bolhuis_transition_2002}. The maximum length of a shifting move was 80~ps. After this equilibration a further 7500 TPS moves were performed as production run.

To classify a molecule $i$ as either mobile or immobile we employed the {\it enduring displacement} formalism~\cite{speck_constrained_2012}:
\begin{equation}
m_i(t)=h(|\mathbf{\bar{r}}_i(t+\Delta t)-\mathbf{\bar{r}}_i(t)|-a),
\end{equation}
where $a = $ 1\r{A}, $\Delta t = $ 2~ps, and $\mathbf{\bar{r}}_i(t)$ is the inherent structure position of molecule $i$ at time $t$ and the Heaviside step function $h(x)$. The parameters $a$ and $\Delta t$ were chosen to probe the nearest neighbor environment on typical time scales where molecules either jump from or remain close to their original position. We also calculated the $\chi_4$ for mW at 235~K and find that $t_0$ lies between 1 and 4~ps. To find the inherent structure positions, the FIRE algorithm was used~\cite{bitzek_structural_2006}. The coarse-grained immobility ($\mathcal I$) and crystallinity ($\mathcal Q$) fields are then defined as follows:
\begin{align}
  \mathcal{I}(\mathbf{r}) &= \sum_{i=1}^{N}(-1)^{m_{i}}  G(|\mathbf{r}-\mathbf{r}_{i}|), \text{ and } \\
  \mathcal{Q}(\mathbf{r}) &= \sum_{i=1}^{N}(-1)^{q_{i}-1} G(|\mathbf{r}-\mathbf{r}_{i}|)
\end{align}
\noindent where the sum runs over all molecules, $G(r) = (2\pi\xi^{2})^{-\frac{3}{2}}\exp\left(-r^{2}/2\xi^{2}\right)$ is a normalized Gaussian with $\xi=2.8$\,\AA. 
$m_{i} = 1$ and $q_{i}=1$ if the molecule identified with the position vector $\mathbf{r}_{i}$ is {\it mobile} and {\it ice-like} respectively. 

\begin{acknowledgments}
We thank L. Joly, A. Zen, B. Slater, C. G. Salzmann, D. Limmer and D. Chandler for stimulating discussions and suggestions. This work was supported by the ERC under the European Union's Seventh Framework Programme (FP/2007-2013)/ERC Grant Agreement 616121 (HeteroIce project). AM's initial work on this project was supported by the Miller foundation at UC Berkeley. We acknowledge the use of the UCL Grace and Legion facilities, the ARCHER UK National Supercomputing Service through the Materials Chemistry Consortium through the EPSRC Grant No. EP/L000202 and the UK Materials and Molecular Modelling Hub, which is partially funded by EPSRC (EP/P020194/1).
\end{acknowledgments}

\bibliography{Text}

\clearpage
\onecolumngrid 
\setcounter{section}{0}
\renewcommand{\thesection}{S\arabic{section}}%
\setcounter{table}{0}
\renewcommand{\thetable}{S\arabic{table}}%
\setcounter{figure}{0}
\renewcommand{\thefigure}{S\arabic{figure}}%
\section*{Supporting Information}
\raggedbottom

\section{The Role of Statistics for calculating the DP}
\begin{figure}[ht]
	\centerline{\includegraphics[width=11cm]{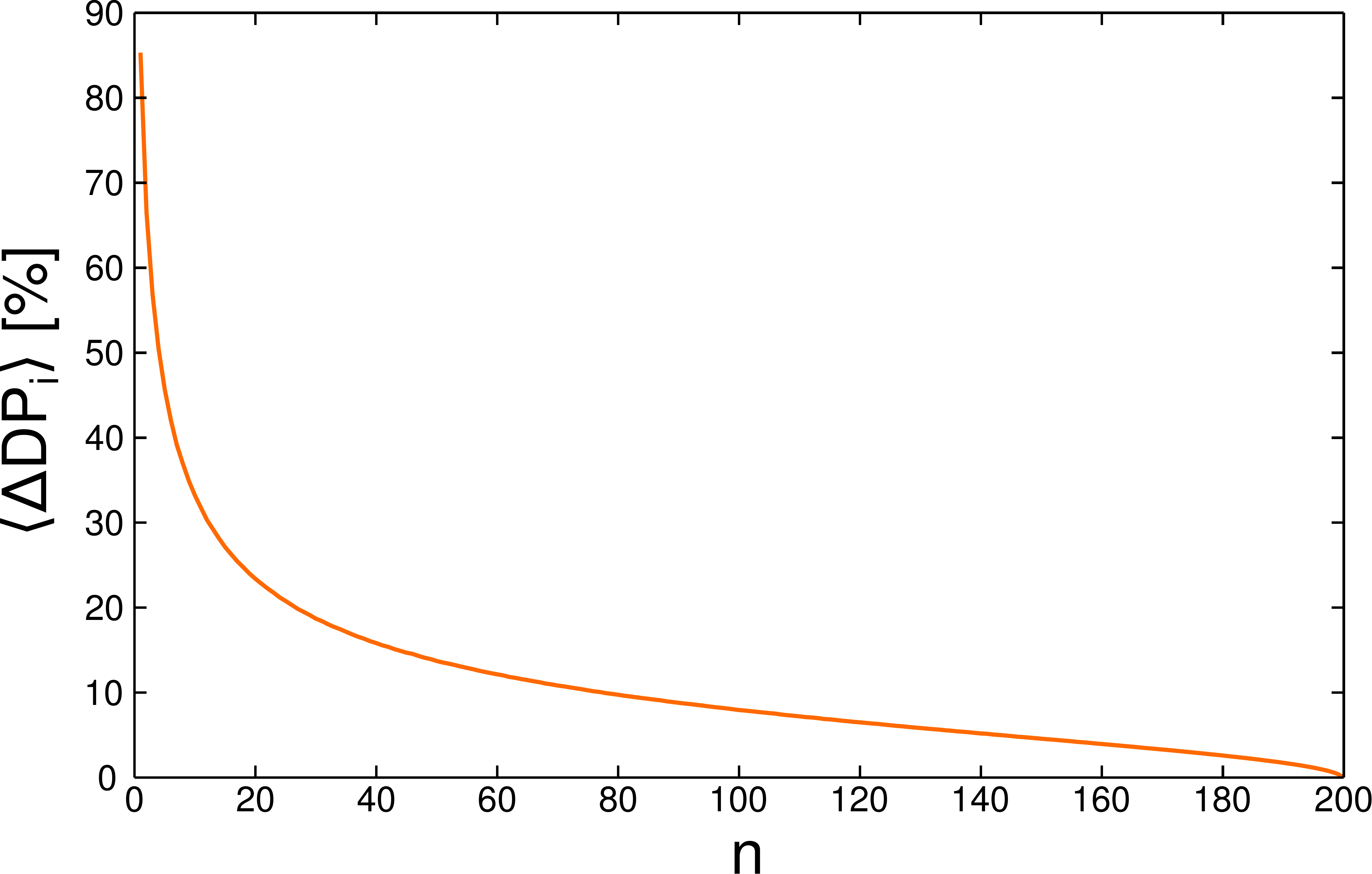}}
	\caption{Assessment of the role of statistics on the calculation of dynamical propensity (DP). Shown is the average difference (in \%) of the calculated DP value for each molecule, when using $n$ instead of 200 runs to calculate the square-displacement. The estimate was done with Jackknife re-sampling~\cite{efron_jackknife_1981}.}
	\label{FIG_S1}
\end{figure}
The growing time of maximum heterogeneity $t_0$ with increasing supercooling also increases the cost of the iso-configurational analysis. Therefore we aim at choosing a number of shootings $n$ for each snapshot that is a compromise between a reasonable accuracy of the dynamical propensity and computational saving. In figure~\ref{FIG_S1} we show an assessment of the influence of choice of $n$. We estimated the average difference of all DP values in a given snapshot obtained with $n$ shootings compared to the estimate with 200 shootings, where we used the latter number to characterize snapshots for the rings-analysis. We see from this that even halving the number of runs to $n=100$ would introduce an average error of only less than 10\% on the DP values which is not enough to qualitatively change any of our conclusions.

For certain parts of this work many frames had to be analyzed regarding their DP distribution. This is a substantial computational effort, especially for the atomistic model which has a time of maximum heterogeneity $t_0$ (which we set as length of each run) of more than 100~ps at 240~K and more than 600~ps at 230~K. Thus we reduced the number of runs for the ISOCA from 200 to 40 for (i) the overlap calculation of regions with ice-like clusters (Fig. 2b in the main text); (ii) the plot of the temporal evolution of the DP before cluster assembly (Fig 2c in the main text); and (iii) the supplementary video. As we can see in figure~\ref{FIG_S1}, large average deviations for the DP values are only obtained for choices of $n < 20$. For $n=40$ we obtain an average deviation of 15\%. We can see that doubling $n$ to 80 would marginally increase the accuracy while also doubling the cost for simulation. Furthermore, an average deviation of 15\% is unlikely to change our results regarding the structural analysis of the resulting domains since this is not enough to significantly change the ranking of the top/bottom 5\% of DP values.

\clearpage
\section{DP Threshold Choice for Defining MI and MM Domains}
\begin{figure}[ht]
	\centerline{\includegraphics[width=12cm]{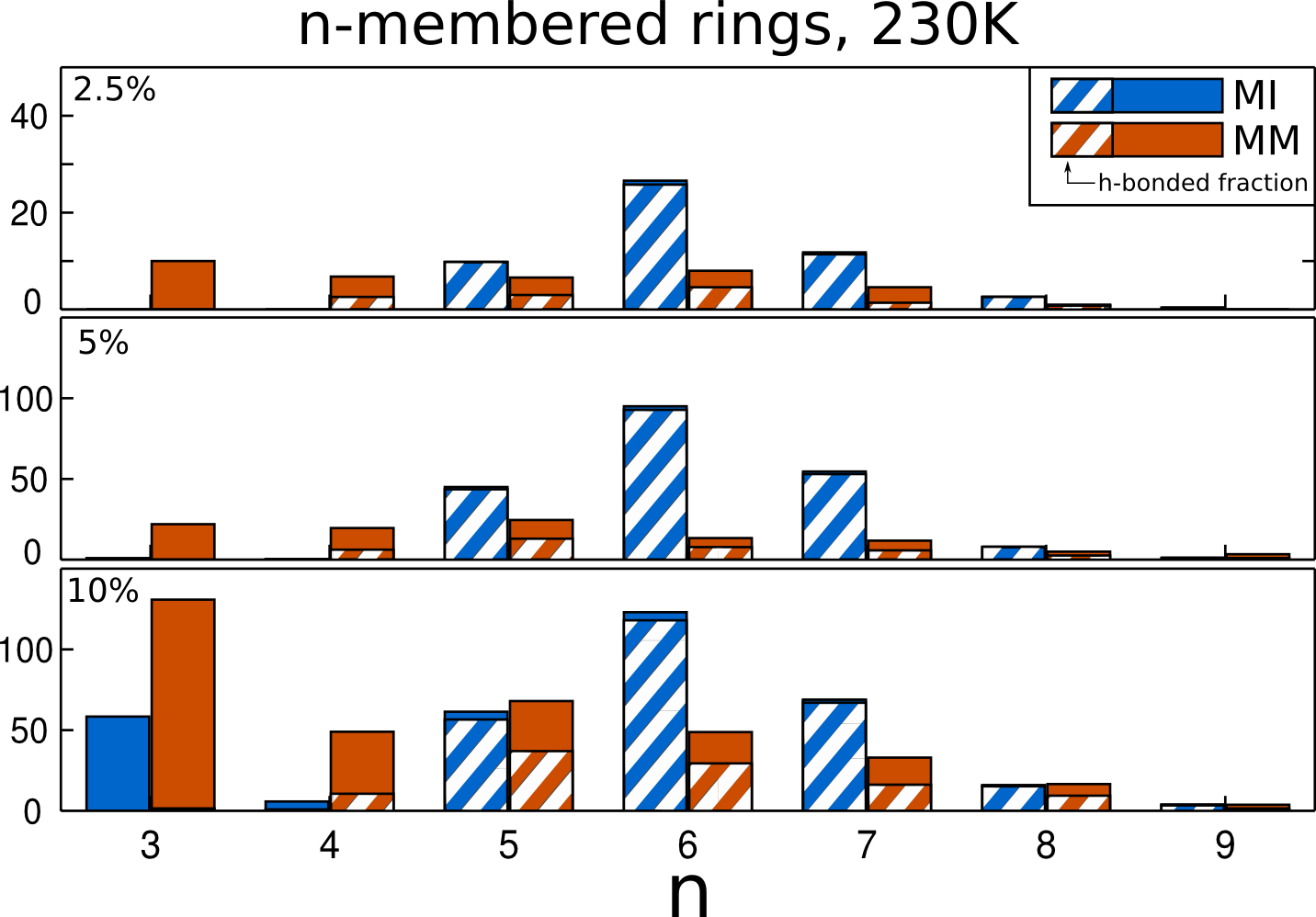}}
	\caption{Number of n-membered rings found in a system of 10000 TIP4P/ICE water molecules at 230~K. The three plots show results for using the top/bottom 2.5\%, 5\% and 10\% to characterize mobile/immobile regions.}
	\label{FIG_S2}
\end{figure}
To investigate the structural characteristics of MI and MM domains we make a DP threshold choice $\epsilon$ for the fraction of molecules we will label as mobile/immobile if they belong to the top/bottom $\epsilon$-fraction of DP values. This choice is in principle arbitrary and since in the main text we reported results for a choice of $\epsilon=$5\%, we here show the results of the rings-analysis for halving/doubling this threshold. In figure~\ref{FIG_S2} we can clearly see that the characteristic occurrence of 6-membered rings for the MI domain and of smaller / less hydrogen-bonded rings in the MM region is not altered by that choice. This suggests that our findings are truly characteristic structural signatures of the most mobile / most immobile molecules in supercooled water.

\clearpage
\section{Correspondence between DP and inherent structure displacements}
\begin{figure}[ht]
	\centerline{\includegraphics[width=11cm]{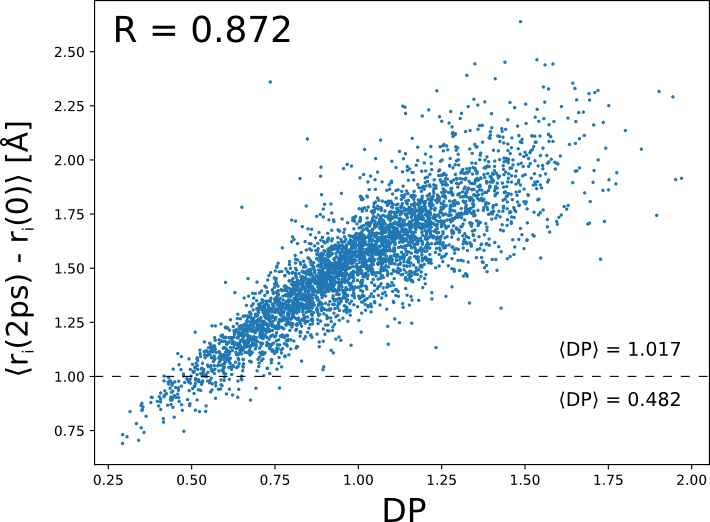}}
	\caption{Average inherent structure displacement versus dynamical propensity of each
molecule in a system containing 4000 mW molecules at 235~K. The two DP averages are the average DP of molecules that have inherent structure displacements below or above 1\r{A} (dashed line, same cutoff as used in the main article).}
	\label{FIG_S9}
\end{figure}
The two methods used in the main article to quantify the mobility of each molecule differ from each other in that the DP formalism averages out the velocities while the enduring displacement formalism takes a particular set of positions and momenta. The latter method is well suited to study our TPS ensemble as the analysis can be done on the fly and calculating the DP at each frame of each trajectory would be much more costly. For the atomistic model we utilized the DP formalism as we were interested in the effects of structure alone, specifically the H-bond network, and the whole workflow is also established in the literature~\cite{sosso_dynamical_2014}. To be able to make a fair comparison between the two it is necessary to either study the DP without averaging velocities or calculate the inherent structure displacements used in the enduring displacements formalism for many different starting velocities. Here, we do the latter for simplicity to study how well the two methods correspond to each other.

We calculated for a mW system containing 4000 molecules the average displacements between the inherent structures, starting from the same initial snapshot 1000 times, with a lag time of 2~ps. All settings were consistent with what was done in the TPS part of the main article. With the same time lag time and number of runs we have established the DP of the molecules in this system. We compare the results in Fig.~\ref{FIG_S9}, which shows the inherent structure displacement versus the DP of each molecule. In there, it can clearly be seen that there is a strong correlation between these two quantities (also indicated by a high Pearson-correlation coefficient $R$ > 0.8). Furthermore, if we consider molecules that would be classified as immobile by enduring displacements (if their inherent structure displacement is below 1\r{A}, see main article formula 4), which is the area below the dashed line in Fig.~\ref{FIG_S9}, we find that their DP is substantially lower ($\sim$0.482) than the average of 1. From these results, it is therefore reasonable that we should not expect too much difference whether we use enduring displacements with inherent structures, or DP from ISOCA, even though we can acknowledge that the two methods are not strictly equivalent. Rather, they relate to the same core physical character of a molecule, which is its mobility and the central aim of this paper. We note here explicitly, that the initial velocities for the two types of simulations were not the same, i.e. the two quantities in Fig.~\ref{FIG_S9} were calculated from different molecular dynamics runs. The fact that we still get such a high correlation is even more encouraging, as the expected correlation between the two different methods is expected to be even higher if they are calculated from the same simulations.

\clearpage
\section{Domain Characterization With Other Order Parameters}
\begin{figure}[ht]
	\centerline{\includegraphics[width=13cm]{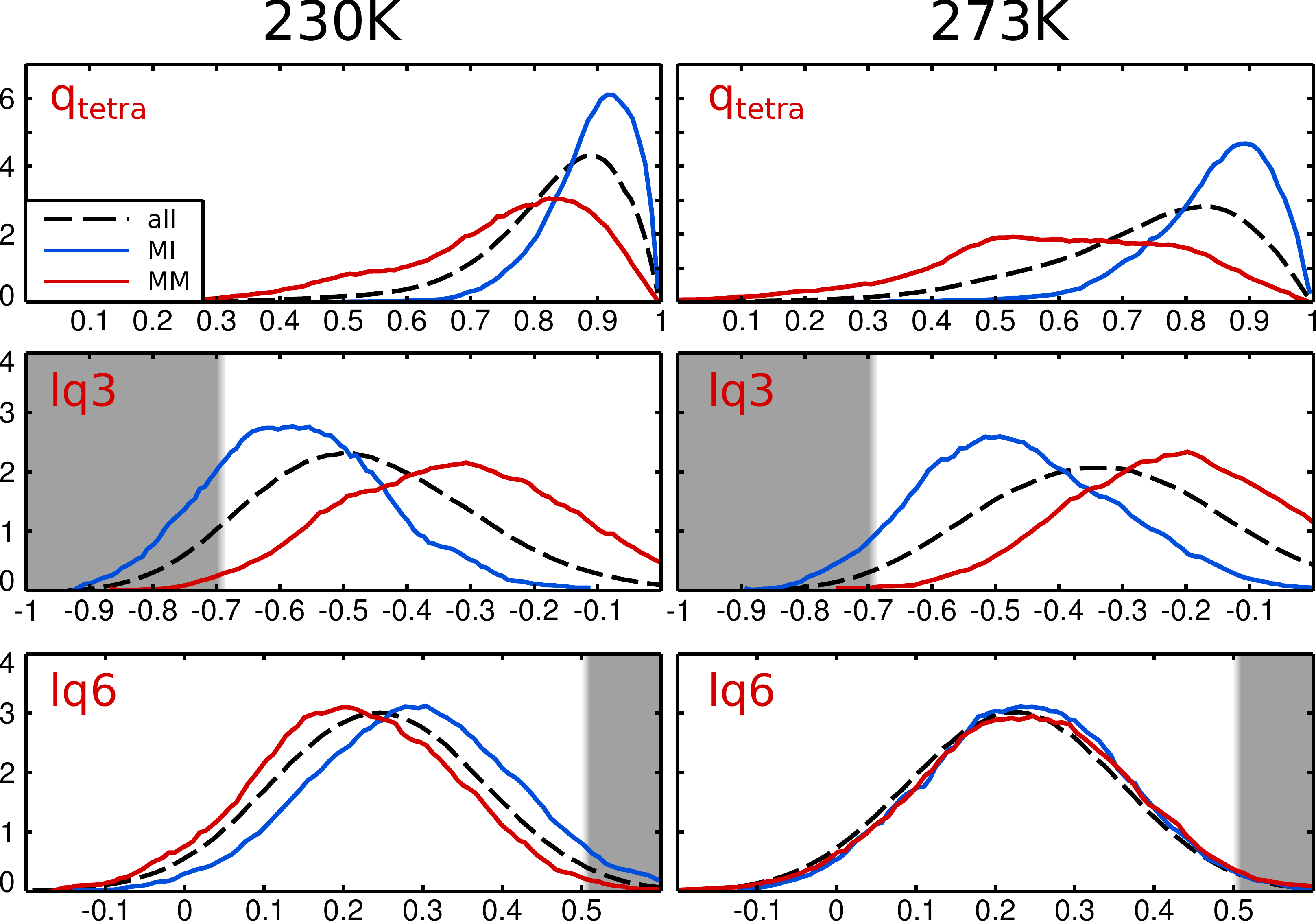}}
	\caption{Characterization of different domains by distributions of customarily used order parameters: the tetrahedrality measure for each molecule q$_\mathrm{tetra}$~\cite{errington_cooperative_2002} and phase-averaged local Steinhardt parameters~\cite{steinhardt_bond-orientational_1983,li_homogeneous_2011}, using spherical harmonics of order 3 and 6 (lq3 and lq6). lines for the most immobile (MI) / most mobile (MM) region are blue/red while averages over the whole simulation cell are black. The gray shaded regions show values of the order parameter that indicate an ice-like environment.}
	\label{FIG_S3}
\end{figure}
After having identified the MI and MM domains we performed comprehensive structural analysis in terms of rings population and hydrogen bonding. In figure~\ref{FIG_S3} we also report the distributions for other order parameters for the aforementioned regions as well as in the overall snapshot. The measures for tetrahedrality $q_\mathrm{tetra}$ and local structuring lq3 and lq6 all show minor differences between the regions. The most important result from this is the fact that the overlap with the ice-like region of the respective order parameter (indicated in gray) is marginal, suggesting that a drop in mobility could precede nucleation. 

We also calculated the number of H-bonds per water molecule according to the simple $R$-$\beta$ criterion outlined in Ref.~\citenum{kumar2007hydrogen}. The average relative populations of molecules with 1, 2, 3, 4, 5 H-bonds at 230~K are 0.02\%, 0.64\%, 8.53\%, 88.70\% and 2.11\%, which is in good agreement with Gasparotto et al.~\cite{gasparotto2016probing}. The almost 89\% of the population that have 4 H-bonds indicates a lack of direct spatial correspondence with either MI or MM regions. In other words, the spatial density of H-bonds does not seem to be extremely heterogeneous since almost all molecules have 4 H-bonds. We also computed the average DP of the H-bond populations, resulting in 1.765, 1.561, 1.281, 0.962 and 1.283. Since the average DP across all molecules is 1 by definition this clearly shows that there is a (very slight) tendency for water molecules with exactly 4 H-bonds to be more immobile, while all other H-bond configurations tend to increase the mobility.

Lastly we note that an analysis of the MI and MM domains in terms of hexagonal and double-diamond cages (see e.g. Refs.\citenum{haji-akbari_direct_2015,sosso_microscopic_2016}) did not yield any significant occurrence of such ($\sim 1$ per snapshot), further indicating that our configurations even at the strongest supercooling are still liquid.

\clearpage
\section{Domain Characterization for the mW Model}
\begin{figure}[ht]
	\centerline{\includegraphics[width=14cm]{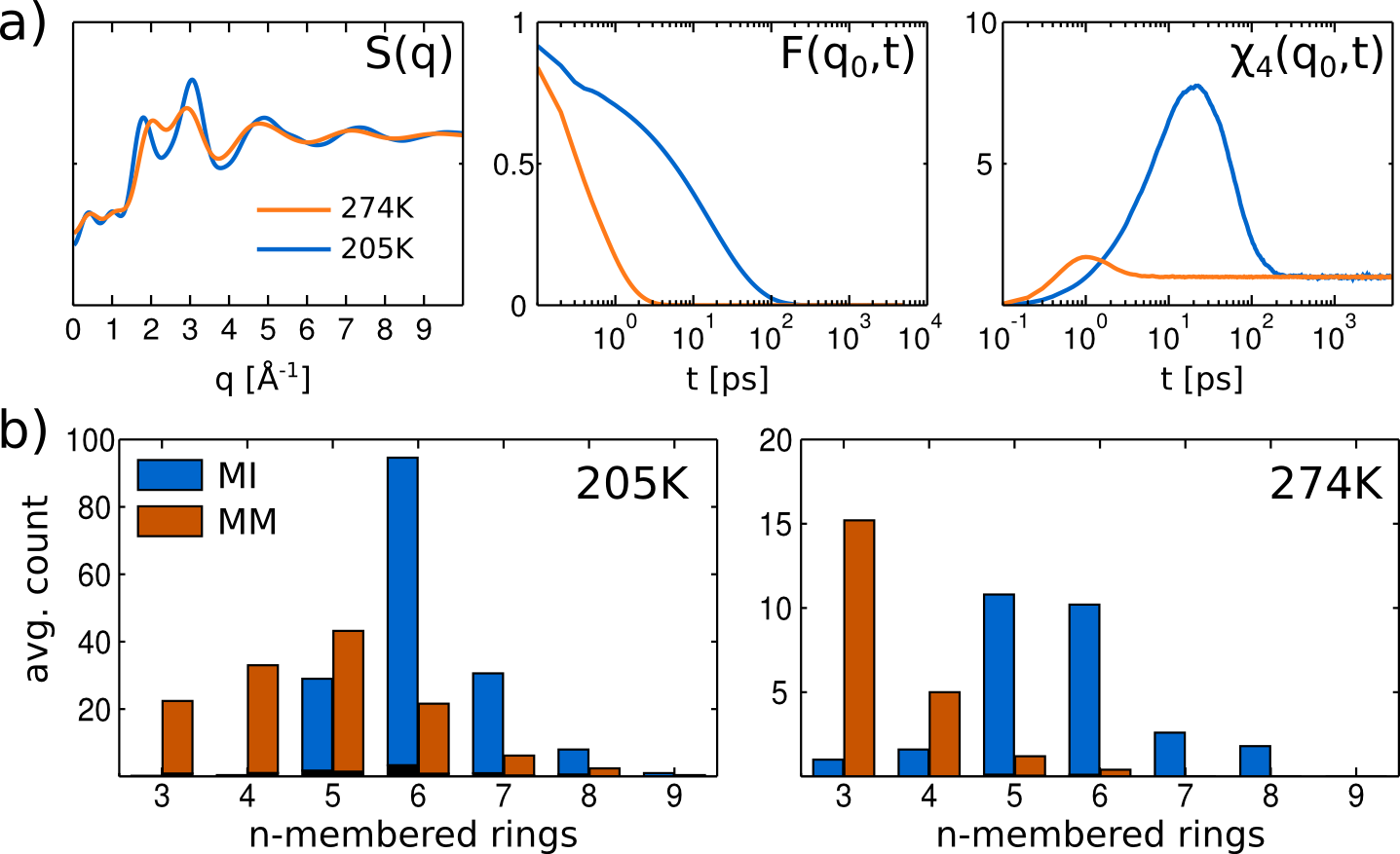}}
	\caption{Dynamical heterogeneity for the mW model~\cite{molinero_water_2009}. a) Structure factor $S(q)$, self-intermediate scattering function $F(q_0,t)$ and dynamical susceptibility $\chi_4(q_0,t)$ at the melting point (274~K) and slightly above the spinodal region (205~K). b) Average number of n-membered rings at the two temperatures in a homogeneous system of 10,000 molecules. The black portion (barely visible) of the bars represents the ice-like fraction of molecules in the respective rings. Despite the fact that the coarse-grained mW model shows much less dynamical heterogeneity the population of rings in MI and MM regions is very similar to the results for the fully atomistic model.}
	\label{FIG_S4}
\end{figure}
The entire workflow performed mostly on the all-atom model was repeated for the mW model at coexistence (274~K) and close to the homogeneous freezing temperature (205~K). In figure~\ref{FIG_S4} we see that the extent of dynamical heterogeneity is much weaker than for the all-atom model. We also calculated the $\chi_4$ for mW at 235~K and find that $t_0$ lies between 1 and 4 ps, indicating that this choice of 2~ps as lag time for the TPS analysis is suitable to distinguish heterogeneity of movements. Only at the low temperature (which corresponds to a supercooling of $\approx 70$~K) there is a hint of heterogeneous dynamics comparable to that of the other model for $\approx 10$~K supercooling. This is expected and in agreement with the fact that this liquid has a larger diffusion coefficient~\cite{molinero_water_2009}. However, the population of the rings network shows similar signatures regarding the type and amount of rings that is seen in the MI and MM regions. We take this as support that results for the mW liquid are still indicative for real water.

\clearpage
\section{Transition Path Sampling for the mW Model}
\subsection{Statistical Independence of Trajectories}
\begin{figure}[ht]
  \centerline{\includegraphics[width=0.62\linewidth]{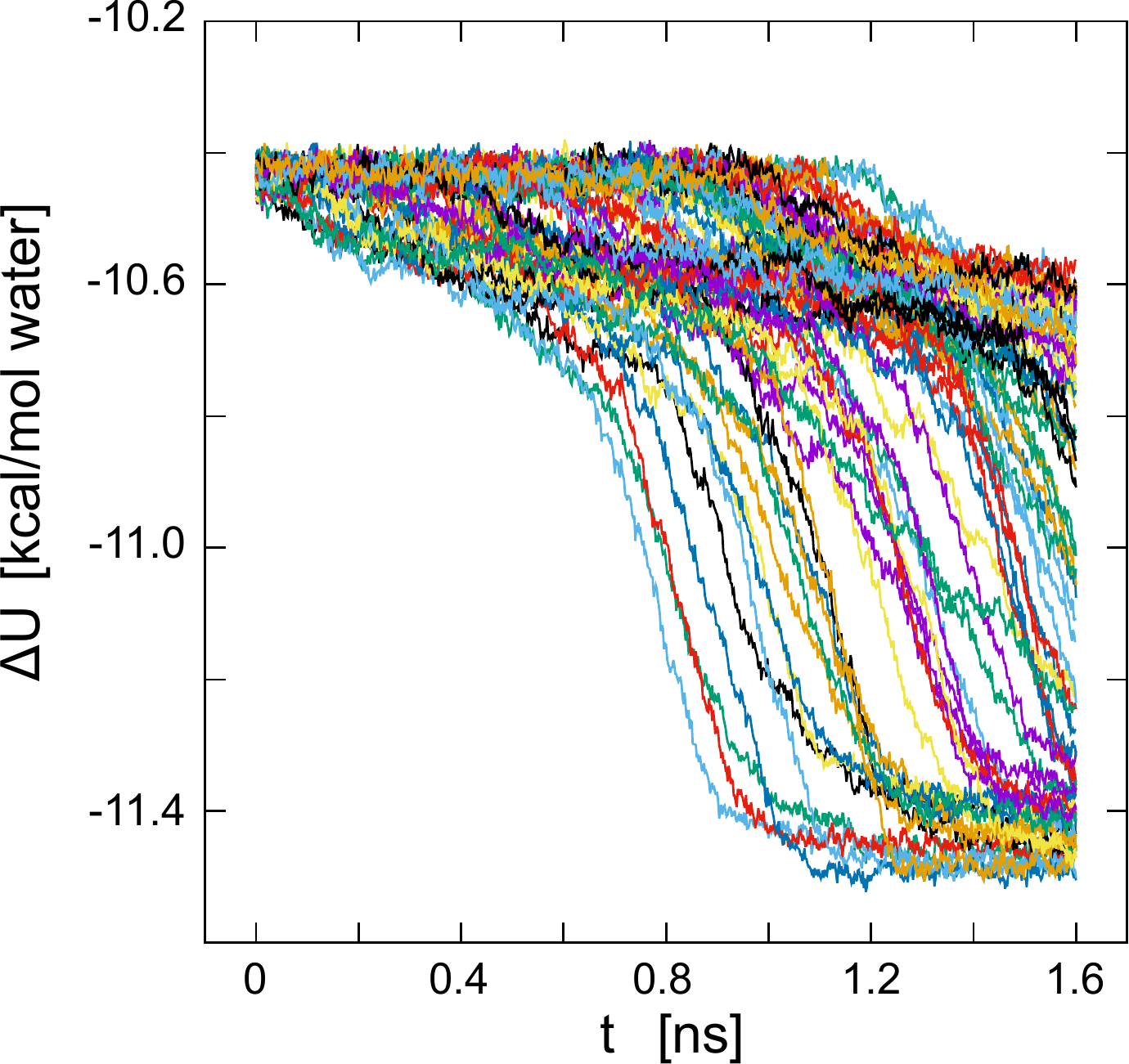}}
  \caption{Potential energy versus time for the trajectories harvested with TPS.}
  \label{FIG_S5}
\end{figure}

When performing a transition path sampling (TPS) study we need to address whether or not we are effectively sampling different trajectories. In Fig.~\ref{FIG_S5} we show the potential energy vs time for all of the trajectories harvested using TPS (76 trajectories from a total of 7500 TPS moves, saving every 100 trajectories.) The plot clearly illustrates that many types of trajectory are being sampled: Some have long induction times while others look like they start nucleating straightaway; others reach a much lower potential energy state than others too. It also appears that there is a range of crossing times. Perhaps most importantly, the trajectories go back-and-forth between these different types of trajectory, suggesting that we are actually sampling a steady state distribution in trajectory space rather than still relaxing the trajectories toward some steady state. 

\clearpage
\begin{figure}[ht]
  \centerline{\includegraphics[width=0.68\linewidth]{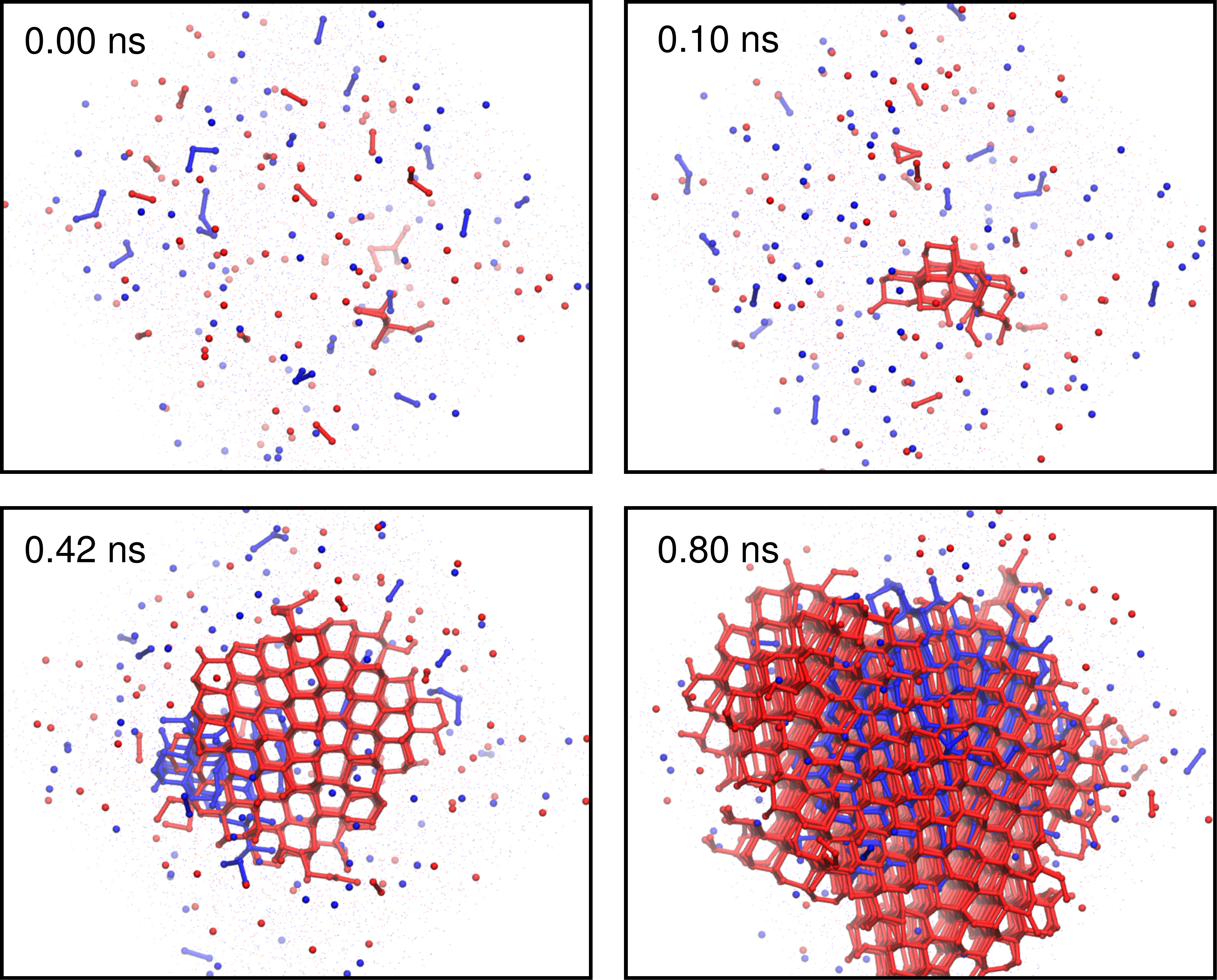}}
  \caption{Indicative snapshots from trajectories that differ by 100 TPS moves.}
  \label{FIG_S6}
\end{figure}
Fig.~\ref{FIG_S6} shows snapshots from two trajectories that differ by 100 TPS moves. Ice-like molecules, with lq$_{6}>0.5$, are shown in red and blue for the different trajectories. The simulation cells were aligned at their fixed NPT point and have the exact same orientation. It is clear that the red and blue clusters (i) originate in different locations (see panels for 0.10~ns and 0.42~ns); (ii) originate at different times; and (iii) have a different orientation with respect to each other (see panel 0.80~ns). This shows that these two examples of trajectories are decorrelated and that our TPS sampling works as intended.

In order to quantify the statistical independence of the trajectories, we have considered the following quantity for trajectory $m$:
\begin{equation}
  \mathcal{N}_{m} = \sum_{i\in \text{trans}} N_{\text{core}}(t_{i})
\end{equation}
\noindent The notation ``$i\in\text{trans}$'' indicates that we are only summing over configurations for which $10<N_{\text{core}}<500$. $N_\mathrm{core}$ was calculated according to the clustering criterion outlined in the Methods section of the paper, i.e. it includes all molecules that are ice-like (lq$_{6}>0.5$) and within 3.4~\r{A} of each other. For the order parameter $N_\mathrm{cls}$ (used to decide of a trajectory reached a basin) we also included surface molecules, i.e. each ice-like molecule's nearest-neighbor, into the cluster so that $N_\mathrm{cls} = N_\mathrm{core} + N_\mathrm{surf}$~\cite{li_homogeneous_2011}. The decay of the autocorrelation function $\propto \left\langle \left(\mathcal{N}_{0} - \langle \mathcal{N} \rangle \right)\left(\mathcal{N}_{m} - \langle \mathcal{N} \rangle \right)\right\rangle$ will then give an indication of how correlated the trajectories are. The correlation ``time'' from this analysis is $\tau \approx 75$\,trajectories, indicating that every 100$^{\text{th}}$ trajectory should be close to statistically independent. 

\clearpage
\subsection{Coarse-Grained Representation}
\begin{figure}[ht]
  \centerline{\includegraphics[width=0.42\linewidth]{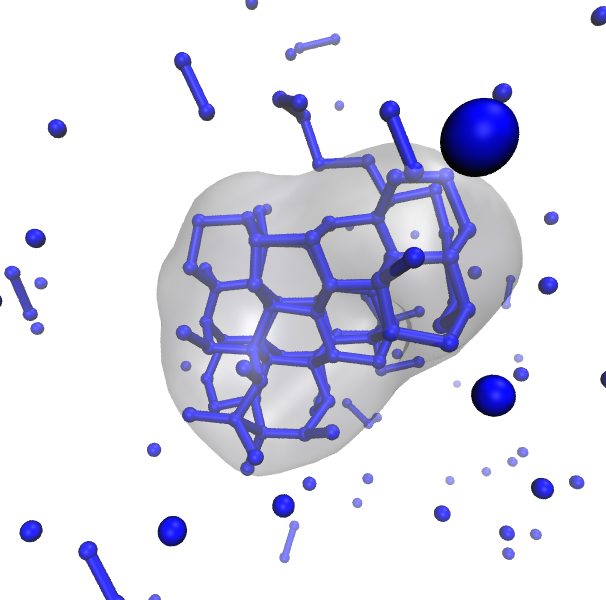}}
  \caption{Verification of the coarse-grained representation of ice-like space. In grey we show the coarse-grained field for $\mathcal{Q}(\mathbf{r})>0$, while blue spheres and bonds correspond to ice-like molecules characterized with the lq$_6$.}
  \label{FIG_S7}
\end{figure}

In the main text we introduced the coarse grained fields $\mathcal{I}$ and $\mathcal{Q}$ to analyze the trajectories harvested from TPS. We note that the framework we have used to compute $\mathcal{I}$ and $\mathcal{Q}$ bears a strong resemblance to that of Willard and Chandler~\cite{willard2010instantaneous}, although the determination of interfaces is simplified in the current context. To demonstrate the method, Fig.~\ref{FIG_S7} shows the `blue trajectory' from Fig.~\ref{FIG_S6} at 0.42\,ns. The transparent iso-surface is representing regions of space where $\mathcal{Q}(\mathbf{r})>0$. In practice, $\mathcal{Q}$ and $\mathcal{I}$ are evaluated on a $20 \times 20 \times 20$ grid, which corresponds to a grid spacing of $\sim 2.5$~\r{A} (which fluctuates owing to NPT dynamics). We can see that this coarse graining procedure does a very good job of encompassing the nascent ice nucleus. Due to the fact that ice-like and liquid-like regions contribute with different signs, this procedure also benefits from the fact that only regions of space that are very ice-like appear to give $\mathcal{Q}(\mathbf{r})>0$: We can see that small clusters away from the nucleus do not show up in this coarse grained representation.

\subsection{Ensemble Connection Between Structure and Dynamics}
In the main article we presented results from a single representative trajectory harvested from TPS. In order to glean information from the entire ensemble of trajectories obtained we introduce:
\begin{eqnarray}
   s_\mathrm{ice/im} &=& \frac{1}{N_\mathrm{ice}}\sum_{i\in
     \mathrm{ice}}\mathrm{sgn}\big(\mathcal{I}(\mathbf{r}_{i})\mathcal{Q}(\mathbf{r}_{i})\big) \label{eqn:iceim} \\
   s_\mathrm{liq/mo} &=& \frac{1}{N_\mathrm{liq}}\sum_{i\in
     \mathrm{liq}}\mathrm{sgn}\big(\mathcal{I}(\mathbf{r}_{i})\mathcal{Q}(\mathbf{r}_{i})\big), \label{eqn:liqmo}
\end{eqnarray}
where the sum over $i$ in Eq.~\ref{eqn:iceim} runs over $N_\mathrm{ice}$ ice-like ($\mathcal{Q}(\mathbf{r}_{i}) > 0$) grid points, while in Eq.~\ref{eqn:liqmo} we sum over all $N_\mathrm{liq}$ liquid-like grid points ($\mathcal{Q}(\mathbf{r}_{i}) < 0$). Both $s_\mathrm{ice/im}$ and $s_\mathrm{liq/mo}$ are measures of correlation between structure and mobility: If each ice-like region is also immobile, then $s_\mathrm{ice/im} = +1$ (conversely, if each ice-like region is mobile then $s_\mathrm{ice/im} = -1$); similarly, $s_\mathrm{liq/mo} = +1$ if each liquid-like region is mobile, and $s_\mathrm{liq/mo} = -1$ if liquid-like region is immobile. We can average these quantities over all TPS trajectories, restricting either to configurations containing an ice-like cluster (with $100 < N_\mathrm{cls} < 200$, denoted as $\langle\cdot\rangle_\mathrm{cls}$), or configurations that are liquid like ($N_\mathrm{cls} < 15$, denoted as $\langle\cdot\rangle_\mathrm{liq}$).
Following this procedure, we first find that $\langle s_\mathrm{ice/im}\rangle_\mathrm{cls} = +0.997\,\pm\,0.001$ which indicates that ice-like regions and immobile regions are highly correlated for cluster sizes relevant to nucleation. We have noticed that the ice nucleus appears to be surrounded by a non-crystalline immobile region. Indeed, we find that $\langle s_\mathrm{liq/mo}\rangle_\mathrm{cls} = +0.58\,\pm\,0.01$, significantly lower than $\langle s_\mathrm{liq/mo}\rangle_\mathrm{liq} = +0.773\,\pm\,0.005$, indicating that the nascent ice-nucleus also reduces the mobility of the surrounding liquid (see also section~\ref{sec.arrested}).

\clearpage
\section{Arrested Dynamics Around Ice Clusters}
\label{sec.arrested}
\begin{figure}[ht!]
\begin{centering}
\centerline{\includegraphics[width=10cm]{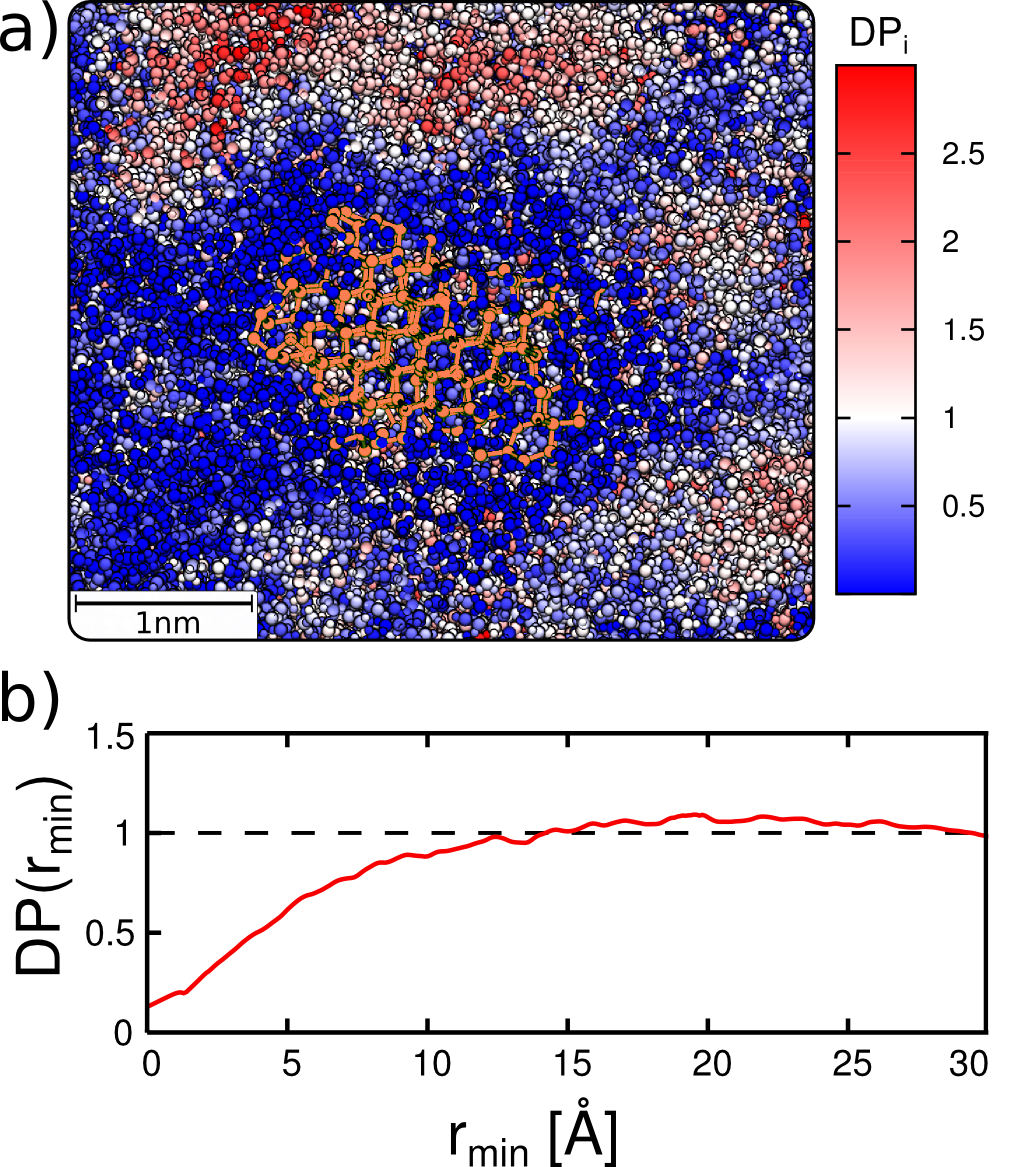}}
\par\end{centering}
\protect
\caption{Dynamical heterogeneity (DH) in presence of an ice-cluster at 230~K. a) DH of water molecules (spheres colored according to the color scale) surrounding a critical ice-cluster (orange bonds and spheres). The presence of the cluster causes a ``cloud'' of immobility around it. b) Average DP of a water molecule as a function of its minimum distance to the cluster $r_\mathrm{min}$. The slowdown of dynamics around the cluster extends for $\sim$10~\r{A}, which roughly corresponds to 2.5 hydration layers.}
\label{FIG_S8}
\end{figure}

In our work we have identified that ice nucleation occurs in relatively immobile regions. To complete the emerging picture we further studied the effect of a sizeable ice-cluster on the dynamics of the surrounding water with the TIP4P/Ice model. We considered an ice cluster from an earlier study~\citep{sosso_ice_2016} that consists of $\sim$ 300 water molecules, which is roughly the critical cluster size at 230~K~\citep{haji-akbari_direct_2015}. The ice cluster is solvated with 100,000 water molecules in a large cubic box (side length $\sim$16~nm) to avoid finite-size effects in the DP and cluster self-interaction. After equilibrating the simulation box at 300~K, whilst keeping the cluster fixed, we perform a quench to 230~K. Then we run several molecular dynamics steps (for roughly two times the relaxation time of TIP4P/Ice at 230~K each), alternating between keeping the cluster and the water molecules fixed. In this manner we equilibrate the interface between the two, avoiding that the cluster grows or shrinks significantly. After equilibrating the interface, we performed an ISOCA on the resulting system, incorporating only liquid molecules into the calculation of the DP. In Fig.~\ref{FIG_S8}a we show a slice through the ice-cluster (orange) which we find to be surrounded by a ``cloud'' of immobile molecules (that are not part of the ice but rather the liquid). The resulting MI regions are of comparable size to the cluster itself. This indicates that ice-clusters not only prefer to appear in immobile regions, but once a large cluster has formed it further slows down the surrounding water. As shown in the previous section the TPS results also show a surrounding shell of immobile particles around sizeable ice-like regions. This is relevant to crystal growth as the liquid molecules in direct vicinity are significantly less mobile than those in the bulk. We have quantified the range of the cluster influence in Fig.~\ref{FIG_S8}b where we see that it takes approximately 10~\r{A} ($\sim 2.5$ hydration layers) for the dynamics to return to the average. This is roughly the same distance that was found to mark the change from confined/interfacial to bulk behavior of water in nanopores~\cite{klameth_structure_2013} and at solid surfaces~\cite{bjorneholm2016water}, indicating an interesting correspondence between the change in dynamics via confinement and vicinity to a crystal.

\clearpage
\section{Video of Nucleation in the Immobile Domain for mW}
We provide the video DH\_mW.mp4 which gives a visual impression of the connection between dynamical heterogeneity and nucleation for a system comprised of 10,000 mW molecules at 205~K. At this temperature the liquid is both dynamically heterogeneous (see figure~\ref{FIG_S8}) and nucleation happens on timescales accessible to standard molecular dynamics. In the representation we show the biggest ice-like cluster in red spheres and bonds as well as the immobile molecules in transparent blue spheres; other water molecules have been omitted for clarity. The immobile molecules have been identified by performing the iso-configurational analysis for each of the $\sim 1800$ frames of the video. Since this involves a series of runs for each of the frames we reduced the number of simulations to $n=40$ compared to $n=200$ for the rings-analysis that had to be done on only 5 frames. In addition we did not normalize by the MSD in the calculation of the DP, i.e.:
\begin{equation}
 \tilde{\mathrm{DP}}_i = \left<\lVert\mathbf{r}_i(t_0)-\mathbf{r}_i(0)\rVert^2\right>_\mathrm{ISO}
\end{equation}
We chose a threshold for $\tilde{\mathrm{DP}}$ for classification of the immobile domains that corresponds to the average bottom 5\% threshold of the DP (calculated with MSD) in the liquid state. This allows for a better visual impression of the dynamics in a system that undergoes a phase-change such as nucleation since the average mobility for the two phases can be very different. In this manner we choose to have on average 500 immobile molecules in the liquid state while there will be more as soon as a sizable cluster forms (since crystalline molecules are less mobile than liquid ones). The video shows that in many frames the biggest ice-like cluster falls into the immobile region and finally the nucleation starts in there as well. After a sizable cluster has formed its surrounding remains permanently immobile.


\end{document}